\documentclass[prd,eqsecnum,10pt]{revtex4}

\usepackage{graphicx}
\usepackage{amsmath}
\usepackage{amssymb}

\newcommand{\be}{\begin{equation}}
\newcommand{\ee}{\end{equation}}
\newcommand{\ba}{\begin{eqnarray}}
\newcommand{\ea}{\end{eqnarray}}

\begin{document}
\title{Critical number of fermions in three-dimensional QED}
\date{\today}
\author{V. P. Gusynin}
\affiliation{Bogolyubov Institute for Theoretical Physics, Kiev, 03680, Ukraine}
\author{P. K. Pyatkovskiy}
\affiliation{Department of Physics and Astronomy, University of Manitoba, Winnipeg, R3T 2N2, Canada}

\begin{abstract}
Previous analytical studies of quantum electrodynamics in $2+1$ dimensions (QED3) have shown the existence
of a critical number of fermions for onset of chiral symmetry breaking, the most known being the value
$N_c\approx3.28$   obtained by Nash to $1/N^2$ order in the $1/N$ expansion [D.~Nash, Phys. Rev. Lett. {\bf62}, 3024 (1989)]. This analysis is reconsidered by solving the Dyson-Schwinger equations for the fermion propagator and the vertex to show that the more accurate gauge-independent value is  $N_c\approx2.85$, which means that the chiral symmetry is dynamically broken for integer
values $N\leq 2$, while for $N\ge3$ the system is in a  chirally symmetric phase. An estimate for the value
of chiral condensate $\langle\bar\psi\psi\rangle$ is given for $N=2$. Knowing precise $N_c$ would be important for comparison between continuum studies and lattice simulations of QED3.
\end{abstract}

\maketitle

\section{Introduction}
Quantum electrodynamics in $2+1$ dimensions (QED3) has attracted much interest over the last three decades. Its parity-invariant version with $N$ flavors of massless four-component Dirac fermions \cite{Pisarski} was extensively used as a test bed for strongly coupled gauge theories as it shares many important features with quantum chromodynamics (QCD) such as confinement and chiral symmetry breaking.
Similar to QCD, in the absence of a bare fermion mass, the model possesses the $U(2N)$ chiral symmetry
which may be broken spontaneously, leading to the dynamical generation of a fermion mass. The main question
that has been debated for a long time is whether chiral symmetry is broken for all values of fermion flavors $N$ or there exists a critical value $N_c$ separating the chiral-symmetric and the chiral-symmetry-broken phases. While at present the majority of works agree on the existence of $N_c$, its precise value remains a matter of debate.

It is remarkable that QED3 has found many applications in condensed matter physics, in particular,
in high-$T_c$ superconductivity \cite{Dorey,superconductivity}, planar antiferromagnets \cite{Farakos},
and graphene \cite{Semenoff} where  quasiparticle excitations have a linear dispersion at low energies
and are described by the massless Dirac equation in $2+1$ dimensions (for graphene, see reviews
in Ref.~[\onlinecite{reviews}]).

Genuine QED3 is ultraviolet finite and has a built-in intrinsic mass scale $\alpha=e^2N/8$
given by the dimensionful gauge coupling $e$,  which plays a role similar to the $\Lambda_{\rm QCD}$
scale parameter in QCD. In the leading order in the $1/N$ expansion, the effective dimensionless coupling,
\be
\bar\alpha(p)=\frac{e^2}{p[1+\Pi(p)]},\quad \Pi(p)=\frac{e^2N}{8 p}, \quad p=\sqrt{p^2},
\label{eff-coupling}
\ee
gives rise to the renormalization-group $\beta$ function
\be
\beta(\bar\alpha)\equiv p\frac{d\bar\alpha(p)}{d p}=-\bar\alpha\left(1-\frac{N}{8}\bar\alpha\right),
\ee
which has the ultraviolet stable fixed point $\bar\alpha=0$ at $p\to \infty$ (asymptotic freedom) and
the infrared (IR) stable fixed point $\bar\alpha=8/N$ at $p=0$. The first study of the Dyson-Schwinger
(DS) equation for the running fermion self-energy in leading order in the $1/N$ expansion has shown \cite{Appelquist} that a phase transition takes place when the infrared fixed point coupling exceeds
some critical value ($8/N>\pi^2/4$). Below the critical number $N_{c}=32/\pi^2\simeq3.24$ the chiral symmetry is broken and a fermion dynamical mass is generated, and above $N_c$ the fermions remain
massless. Hence, the critical number $N_c$ separates the chiral symmetry broken (CSB) phase from the so-called (quasi)conformal phase describing interacting massless fermions and a photon, and the phase transition at $N=N_c$ is supposed to be of infinite order because of the form of the dynamical mass
$m_{\rm dyn}\sim e^2\exp\left(-2\pi/\sqrt{N_c/N -1}\right)$ for $N$ close to $N_c$ \cite{Appelquist,MirYam,Shpagin}. This is similar to what happens in quenched strongly coupled QED4 \cite{QED4-strong,RNC}, where the gauge coupling must exceed a critical value for the dynamical mass generation to occur (note, however, that, in contrast to QED3, the vacuum polarization effects in
QED4 change the infinite order phase transition to the second order one \cite{QED4-polarization}).

The presence of a critical $N_c$ in QED3 is tempting because of possible existence of an analogous
critical fermion number $N_f=N_c$ in (3+1)-dimensional $SU(N_c)$ gauge theories with $N_f$ fermion
flavors \cite{Appelquist-SU(N)}. Also, a nontrivial IR fixed point in QED3 may be related to
nonperturbative dynamics in condensed matter, in particular, dynamics of non-Fermi liquid behavior \cite{superconductivity,Rantner,Kim}.

The analytical approach to study dynamical symmetry breaking in QED3 is based on the self-consistent solution of truncated Dyson-Schwinger equations for the fermion propagator. Numerous papers using
this approach gave the results for the value of $N_c$ in the range $2<N_c<5$
\cite{Appelquist,Nash,GHR-1996,GHR-2001,Maris,Fisher}. Renormalization group studies give the critical
value $N_c$ approximately in the same range \cite{Kubota,Gies} with possible existence of
a third intermediate phase for $N_c<N<N_c^{qc}$ where $N_c^{qc}$ is the ''conformal-critical''
flavor number  above which the theory is in the quasiconformal phase \cite{Gies}. An argument based
on a thermodynamic inequality $f_{\rm IR}\leq f_{\rm UV}$, where $f$ is the thermodynamic free energy
estimated in infrared and ultraviolet regimes, yields the prediction $N_c\leq3/2$ \cite{therm-estimate},
and the most recent bound is $N_c\le4.4$ \cite{Giombi}.

Numerical lattice calculations meet larger uncertainty in determining the critical fermion number
$1<N_c<10$ \cite{lattice}.  Recent paper \cite{Karthik} did not find the chiral condensate
for $N=1,2,3,4$ in contradiction with some previous lattice studies \cite{lattice}. On the other hand,
a numerical study of the quenched ($N = 0$) case has shown that chiral symmetry is broken \cite{Hands}
in accordance with theoretical arguments (see, for example, Ref. \cite{Sizer}). One of the major problems
in studying dynamical symmetry breaking using lattice simulations is the exponential smallness of the
order parameter for $N$ close to $N_c$ and the presence of a massless photon since finite size effects
play a nontrivial role in this case. The presence of an infrared cutoff has been shown to reduce the
value of the critical number of fermions \cite{Reenders-2003}.

The determination of the precise value of $N_c$ is an important task since for many condensed matter
systems the number of four-component Dirac fermions turns out to be $N = 2$, and the system's phase
state depends on whether $N_c$ is above or below two. In the literature, the most cited critical value
is $N_c\approx3.28$ obtained by Nash \cite{Nash} by analyzing the gap equation in the leading and next-to-leading orders of the $1/N$ expansion. Recent paper \cite{Kotikov2016} found the critical
value $N_c = 3.29$, that is only slighter larger than Nash's value. Studies in these two papers were performed in different gauges, the nonlocal gauge with the gauge parameter $\xi=1$ (Feynman-like gauge)
in Ref. \cite{Nash}, and the Landau gauge in Ref. \cite{Kotikov2016}. As we will show, both these
results suffer from gauge dependence of $N_c$, though for different reasons.

The paper is organized as follows. In Sec.~\ref{SDequations}, we formulate our approach for solving
the gap equation for the dynamical mass function. In Secs.~\ref{leading_approx} and
\ref{next-to-leading_approx}, we calculate the critical value of fermion flavors in the leading and
next-to-leading approximations in $1/N$ expansion and show that our value of $N_c$ is gauge independent.
The derived expression for the fermion anomalous mass dimension is in complete agreement with the one obtained by Gracey \cite{Gracey} in the $1/N^2$ order. The numerical estimate of chiral condensate $\langle\bar\psi\psi\rangle$ is obtained in Sec.~\ref{chi-condensate}. The results are summarized and discussed in Sec.~\ref{section-conclusion}. Three Appendixes  are for details
of solving the Dyson-Schwinger equation for the vertex function in three-gamma approximation,
perturbative calculation of the fermion wave-function renormalization, and the Landau-Khalatnikov-Fradkin
transformation for the fermion propagator.

\section{Dyson-Schwinger equations}
\label{SDequations}

\begin{figure}[hpt]
  \begin{center}
    \resizebox{0.55\textwidth}{!}{
    \includegraphics{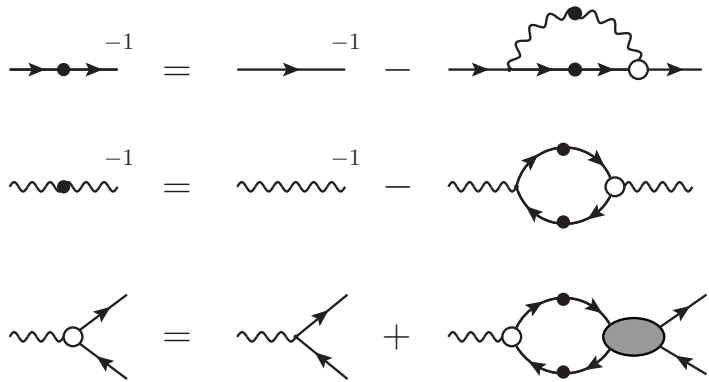}}
    \caption{\label{DSE}The Dyson-Schwinger equations for the
    dressed fermion and photon propagators and the dressed vertex.}
  \end{center}
\end{figure}

We consider QED in three dimensions with $N$ four-component fermion flavors whose
Lagrangian is governed (in Euclidean formulation) by the action
\begin{eqnarray}
S= \int d^3x\left[\bar{\psi}_i\gamma_\mu
D_\mu\psi_i+\frac{1}{4}F_{\mu\nu}^2\right],
\end{eqnarray}
where the covariant derivative $D_\mu=\partial_\mu-ieA_\mu$, $i=1,2,...,N$, and  Euclidean gamma matrices satisfy
$\gamma^\dagger_\mu=\gamma_\mu, \quad
\{\gamma_\mu,\gamma_\nu\}=2\delta_{\mu\nu}.$
The Dyson-Schwinger equations (DSE) for the photon and fermion propagators are given by (see Fig.\ref{DSE})
\ba
\label{fermion:eq}
S^{-1}(p)&=&S_0^{-1}(p)+e^2\int\frac{d^3q}{(2\pi)^3}\,\gamma_\mu
S(q)\Gamma_\nu(q,p) D_{\mu\nu}(q-p),\\
D_{\mu\nu}^{-1}(p)&=&D_{0,\mu\nu}^{-1}(p)-Ne^2\int\frac{d^3q}{(2\pi)^3}\,
{\rm tr}\left[\gamma_\mu S(q)\Gamma_\nu(q,p-q) S(p-q)\right],
\label{photon:eq}
\ea
where $S(p)$ and $D_{\mu\nu}(p)$ are full (dressed) fermion and photon propagators, respectively,
and $\Gamma_\nu(q,p)$ is the full vertex. The vertex $\Gamma_\nu(q,p)$ satisfies its own
Dyson-Schwinger [or Bethe-Salpeter (BS)] equation with the fermion-antifermion scattering kernel
(see Fig.\ref{DSE}).
 The general form of the dressed fermion propagator $S(p)$ and the photon propagator
$D_{\mu\nu}(p)$ is given by
\ba
S(p)&=&\frac{1}{-i\hat{p}A(p)+B(p)},\\
D_{\mu\nu}(p)&=&\left(\delta_{\mu\nu}-(1-\xi)\frac{p_\mu
p_\nu}{p^2}\right)\frac{1}{p^2[1+\Pi(p)]},
\label{photon-prop}
\ea
where the scalar functions $A(p),B(p),\Pi(p)$ depend on $p=\sqrt{p^2}$. In the above equations
$\xi$ is the gauge parameter in a class of covariant nonlocal gauges introduced in Ref. \cite{Nash},
with $\xi=0$ giving the Landau gauge. The functions $A(p),B(p)$  depend on the gauge parameter
$\xi$, on the other hand, the vacuum polarization $\Pi(p)$ is independent of $\xi$.

In what follows, instead of solving DSE for the photon propagator,  we will approximate the vacuum polarization $\Pi$ by its two leading terms of the $1/N$ expansion with massless fermions:
\be
\Pi(p)= \frac{e^2NC}{8p},\quad C=1+\frac{a}{\pi^2N},\quad a= \frac{184}{9}-2\pi^2.
\label{twoloop-Pi}
\ee
For derivation of second term in the constant $C$ see Ref. \cite{GHR-2001}. This expression is valid for
momenta $p<\alpha\equiv e^2N/8$. The mass scale $\alpha$ is kept fixed as $N\to\infty$ and all diagrams
are rapidly damped for momenta $p>\alpha$, which is a reflection of superrenormalizability of QED3. The effective dimensionless coupling given by expression (\ref{eff-coupling}) tends to the infrared fixed
point $8/NC$, thus the dressed photon propagator will be taken precisely at this nontrivial critical point:
\be
D_{\mu\nu}(q)=\frac{8}{e^2NC|q|}P_{\mu\nu}(q),\quad P_{\mu\nu}(q)=\delta_{\mu\nu}-(1-\xi)
\frac{q_\mu q_\nu}{q^2}.
\label{photon-propagator}
\ee

In order to solve the DSE for the fermion propagator one needs to know the fermion-photon vertex
$\Gamma_\nu(q,p)$. In general the vertex satisfies its own DSE which contains an unknown four-point
function, the fermion-antifermion scattering kernel. To avoid complications with solving DSE for the
vertex one usually chooses an appropriate approximation for it, for example the simplest one is the replacement of the full vertex by the bare vertex $\gamma_\nu$ (the ladder approximation). The more sophisticated way is to use some ansatz for the vertex consistent with the Ward-Takahashi identity
\be
i(q-p)_\mu\Gamma_\mu(p,q)=S^{-1}(p)-S^{-1}(q)
\label{vectorWTI}
\ee
and satisfying several other requirements. The Ball-Chiu \cite{BC} and the Curtis-Pennington \cite{CP}
are most known among them. For example, in a paper \cite{Fisher} these ansatze  were used to solve a coupled system of DSE for the photon and fermion propagators and to get an estimate for a critical number
$N_c$ for chiral symmetry breaking in QED3. Although such an approach reproduces the value $N_c\approx4$ close to the value $N_c$ found in leading order of the $1/N$ expansion \cite{Nash,GHR-1996,Maris}, it cannot be considered as a reliable one since the approach does not permit to identify the ansatz with a class of Feynman diagrams. Also one cannot estimate corrections terms to results obtained in this approach.

Taking this into account, in what follows we choose more direct (and standard) way of solving DSE for the fermion mass function. However, instead of solving the equation for the mass function which follows from Eq.~(\ref{fermion:eq}) with one full vertex we will get an equivalent equation written in terms of the
fermion-antifermion forward scattering kernel which is represented by the set of (amputated) two-particle irreducible (2PI) diagrams. This is similar to the approach used earlier in
Refs.~\cite{Baker,RNC,Holdom,Mahanta}.

\begin{figure}[tbp]
  \begin{center}
    \resizebox{0.75\textwidth}{!}{
    \includegraphics{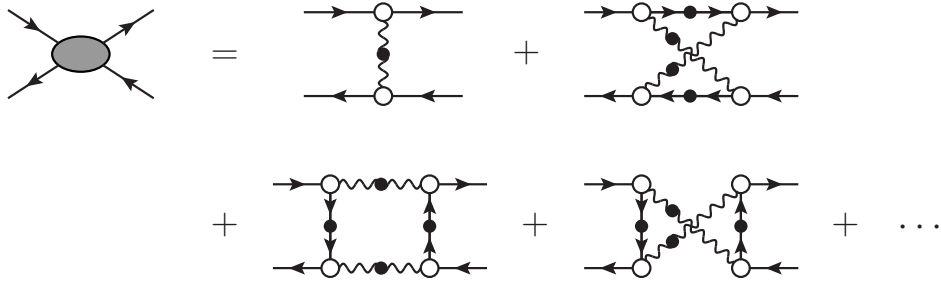}}
    \caption{\label{kernel_skeleton}Skeleton expansion of the fermion-antifermion kernel.}
  \end{center}
\end{figure}

First, we write the BS equation for the axial-vector vertex $\Gamma_{\mu5}$ \cite{Bjorken},
\be
[\Gamma_{\mu5}(p,q)]_{\alpha\beta}=(\gamma_{\mu}\gamma_5)_{\alpha\beta}+\int\frac{d^3k}{(2\pi)^3}
K_{\beta'\alpha';\beta\alpha}(k+q-p,k,k-p)\left[S(k)\Gamma_{\mu5}(k,k+q-p)S(k+q-p)
 \right]_{\alpha^\prime\beta^\prime}.
\label{axialvert:eq}
 \ee
The DSE for $\Gamma_{\mu5}$ is similar to that one for the vertex $\Gamma_{\mu}$ (see Fig.~\ref{DSE})
except for the inhomogeneous term being $\gamma_\mu\gamma_5$ instead of $\gamma_\mu$.
Multiplying the above equation by $i(q-p)_\mu$ and using the axial-vector WTI
\be
i(q-p)_\mu\Gamma_{\mu5}(p,q)=S^{-1}(p)\gamma_5+\gamma_5S^{-1}(q), \label{axialWTI}
\ee
we get in the limit $p\to q$,
\be
(\gamma_5)_{\alpha\beta} B(p)=\int\frac{d^3k}{(2\pi)^3}K_{\beta'\alpha';\beta\alpha}(k,k,k-p)
\left[S(k)\gamma_5B(k)S(k)\right]_{\alpha^\prime\beta^\prime},
\ee
or, in terms of the mass function $\Sigma(p)=B(p)/A(p)$ the last equation can be written as
\be
\Sigma(p)=\int\limits_0^\infty\frac{dkk^2\Sigma(k)}{k^2+\Sigma^2(k)}
K(p,k).
\label{massfunc:eq}
\ee
Here we introduced the notation for the kernel $K(p,k)$:
\be
K(p,k)=\frac{1}{4A(p)A(k)}\int\frac{d\Omega_k}{(2\pi)^3}
(\gamma_5)_{\beta\alpha}K_{\beta'\alpha';\beta\alpha}(k,k,k-p)
(\gamma_5)_{\alpha^\prime\beta^\prime},
\label{K(p,k)_notation}
\ee
where $\int d\Omega_k$ denotes the integration over the angles. The kernel
$K_{\beta'\alpha';\beta\alpha}(k,k,k-p)$ possesses the skeleton expansion with dressed photon
and fermion propagators and full vertices (see Fig.~\ref{kernel_skeleton}). For example, in the
lowest order it is given by the diagram with exchange of one photon
\be
K^{(2)}_{\beta'\alpha';\beta\alpha}(k,k,k-p)=(ie)^2\Gamma_{\mu\alpha\alpha^\prime}(p,k;p-k)
\Gamma_{\nu\beta^\prime\beta}(k,p;k-p)D_{\mu\nu}(p-k).
\label{K_1_photon_exchange}
\ee
The useful feature of Eq.~(\ref{massfunc:eq}) for the mass function is that it does not
contain overlapping diagrams and is multiplicatively renormalizable in contrast to
Eq.~(\ref{fermion:eq}). Eq.~(\ref{massfunc:eq}) is homogeneous in a mass function $\Sigma$,
and as such always has the trivial solution $\Sigma=0$. Our main objective is to find a bifurcation
point $N_c$ where a nontrivial solution bifurcates from the trivial one. Bifurcation theory was first applied to the problem of dynamical mass generation by Atkinson \cite{bifurcation}, and
it remains  one of the main methods used to locate the critical number $N_c$ in QED3 or critical coupling constant in QED4 and QCD (see, for example, Refs.~\cite{Appelquist,Nash,Holdom,Mahanta,ABGRP}).

According to the bifurcation method, in order to find the critical value $N_c$ for the onset of
chiral symmetry breaking we neglect in the kernel of Eq.~(\ref{massfunc:eq}) all terms that are
quadratic or higher in the mass function. Thus we write Eq.~(\ref{massfunc:eq}) in the form
\be
\Sigma(p)=\int\limits_0^\alpha \frac{dk\,k^2\Sigma(k)}{k^2+\Sigma^2(k)} K(p,k),
\label{main:eq}
\ee
where now the kernel $K(p,k)$ is calculated in the massless theory. The integral equation (\ref{massfunc:eq}) is rapidly damped for momenta $p>\alpha$ and the main contribution comes from the
region $p<\alpha$. In the latter region we keep only the lowest-order terms in $p/\alpha$, and put an ultraviolet cutoff $\alpha$. Note that in the massless theory the third and the fourth
diagrams in Fig.~\ref{kernel_skeleton} do not contribute.

It is clear by construction that the kernel $K(p,k)$ is symmetric under exchange $p\leftrightarrow k$.
Next, it can be shown that the quantity $\bar{K}(p,k)=(pk)^{1/2}K(p,k)$ is a function of the ratio
$p/k$. Indeed, in general when external momenta are scaled by $l$, an amputated fermion $n$-point
function gets an anomalous scaling factor $l^{-n\gamma}$ where $\gamma$ is the anomalous dimension
of the fermion field. Because we included the factor $[A(p)A(k)]^{-1}$ into the definition of $K(p,k)$
and $A(p)$ scales as $l^{-2\gamma}$, the anomalous scaling cancels out so that $K(p,k)$ scales according to its canonical dimension $-1$: $K(lp,lk)=l^{-1}K(p,k)$. Since $\bar{K}(p,k)$ is a dimensionless function, it must be a function of $k/p$, $\bar{K}(p,k)\equiv F(k/p)$. It is evident that described symmetry property of $\bar{K}(p,k)$ is independent of the choice of gauge. On the other hand, it is essentially based on the fact that the kernel is considered precisely at the infrared conformal fixed point where the dressed photon propagator behaves according to its canonical scaling $e^2D(p)\sim 1/p$ (this is valid at least in $1/N^2$ order of $1/N$ expansion \cite{GHR-2001}).

Equation~(\ref{main:eq}) is scale invariant except the upper limit of integration. To study the onset of chiral symmetry breaking we look for a powerlike solution $p^{-b}$ of Eq.~(\ref{main:eq}) with infinite upper
limit of integration and neglecting the term $\Sigma^2$ in the denominator, i.e.,
\be
\sqrt{p}\,\Sigma(p)=
\int\limits_0^\infty\frac{dk}{\sqrt{k}}\,\Sigma(k)F(k/p).
\label{scaleinveq}
\ee
It must be emphasized that this is not an approximation but a precise manner to locate the critical point
by applying bifurcation theory \cite{ABGRP}: chiral symmetry breaking occurs when $b$ becomes complex,
which determines the critical value $N_c$ so that the symmetry is broken for $N<N_c$. Equation~(\ref{scaleinveq}) leads to the following equation for the exponent $b$:
\be
1= \int\limits_1^\infty\frac{dx}{x}\left(x^{b-1/2}+x^{1/2-b}\right)F(x),
\label{eq-for-b-exponent}
\ee
where we used the symmetry property of the function $F$: $F(x)=F(1/x)$.
For convergence of the integral
in Eq.~(\ref{eq-for-b-exponent}) the function $F(x)$ should decrease at large $x$ as
$F(x)\propto x^{-\sigma}$ with $\sigma>0$. For example, the order $1/N^2$ calculation
of the kernel in Ref. \cite{Nash} gives $\sigma=1/2+2\gamma$, hence we should assume
$1/2+2\gamma>|b-1/2|$ for convergence of the integral.
Clearly, the critical $N_c$ depends on the level of truncation of the kernel.

\section{The critical $N_c$ in the leading order.}
\label{leading_approx}

Let us see how the above formulae work in case of the simplest approximation (the first diagram
in the skeleton expansion in Fig.~\ref{kernel_skeleton}). In this case it is sufficient to use
the Landau approximation for the full vertex \cite{ Landau} [i.e., $\Gamma_\mu(p,k;q)=
\gamma_\mu A({\rm max}(k,p))$] in Eq.~(\ref{K_1_photon_exchange}) and the equation for $\Sigma(p)$
takes the form
\be
\Sigma(p)=\lambda\int\limits_0^\infty {dk}\frac{\Sigma(k)}{{\rm max}(k,p)}
\frac{A^2({\rm max}(k,p))}{A(p)A(k)},
\ee
where $\lambda=4(2+\xi)/\pi^2N$. For the wave-function renormalization, we take in the leading order
the expression
\be
A(p)\simeq\left(1+\frac{16}{9\pi^2N}\right)\left(\frac{p}{\alpha}\right)^{-2\gamma},
\label{wavefun_renorm}
\ee
where the anomalous dimension $\gamma=2(3\xi-2)/(3\pi^2N)$.
The chosen form for the $A(p)$ function is in agreement with the perturbative calculation in $1/N$,
Eq.~(\ref{right-A-inPT}), and transforms correctly under the Landau-Khalatnikov-Fradkin (LKF)
\cite{LK,Fradkin} transformation between different covariant gauges \cite{Bashir} (see Appendix~\ref{LKF-sec}). The latter property
is crucial for a critical $N_c$ to be gauge invariant. Thus, we have that
\be
F(k,p)=\lambda\frac{(kp)^{1/2}}{{\rm max}(k,p)}\frac{A^2({\rm
max}(k,p))}{A(p)A(k)}
=\lambda\left[\left(\frac{k}{p}\right)^{1/2+2\gamma}
\theta(p-k)+\left(\frac{p}{k}\right)^{1/2+2\gamma}\theta(k-p)\right].
\ee
Then Eq.~(\ref{eq-for-b-exponent}) gives the equation for the exponent $b$:
\be
\Bigl(b-\frac{1}{2}\Bigr)^2=-\lambda(1+4\gamma)+\Bigl(\frac{1}{2}+2\gamma\Bigr)^2\simeq\frac{1}{4}-\frac{32}{3\pi^2N},
\ee
where we kept only terms up to $1/N$ order. Note that the dependence on the gauge parameter $\xi$ has dropped out in the last equation. The exponent $b$ becomes complex for $N<N_c=128/3\pi^2\simeq4.32$
and the onset for complexity determines $N_c$.
We recall that according to the operator product expansion \cite{Cohen}, the parameter $b$ is related to the mass anomalous dimension $\gamma_m$ as $b=1-\gamma_m$ and $\gamma_m$  is a gauge independent quantity; it governs the ultraviolet asymptotic behavior of the fermion dynamical mass function related to spontaneous chiral symmetry breaking: $\Sigma(p)\sim p^{\gamma_m-1}$.

Corrections of the order of $1/N^2$ to the equation for the exponent $b$ in the Feynman-like gauge $\xi=1$ were derived in Ref.~\cite{Nash}, which can be written as (our $b$ differs in sign from $b$ used by Nash)
\be
\Bigl(b-\frac{1}{2}\Bigr)^2=\frac{1}{4}-\frac{32}{3\pi^2N}\Bigl(1-\frac{341+48a}
{48\pi^2N}\Bigr),\quad a=0.706
\label{eq:Nash}
\ee
[Nash's numerical factor $a=0.706$ should be given by our constant $a\simeq0.7052$ defined in Eq.~(\ref{twoloop-Pi})].
The critical $N_c$ is determined from the condition when two roots of this equation become
equal, this happens for $N_c\approx3.28$. For values $N<N_c$, the roots become complex, indicating that oscillatory behavior of the gap function takes over from nonoscillatory one.

Note, however, that Nash's equation (\ref{eq:Nash}) was derived with an error: the number $341$ must
be replaced with $277$, which yields  $N_c=3.52$ instead of the claimed $N_c=3.28$. Besides,
the anomalous dimension of a fermion field in the Feynman-like gauge was calculated with an error in
the order $1/N^2$. This motivated us to reconsider the derivation of the equation for the exponent $b$
in order $1/N^2$.

\section{The critical $N_c$ in the next-to-leading order.}
\label{next-to-leading_approx}
\begin{figure}[tbp]
  \begin{center}
    \resizebox{0.75\textwidth}{!}{
    \includegraphics{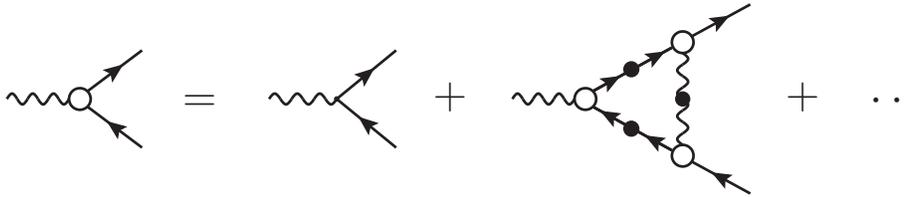}}
    \caption{\label{DSE-vertex}The Dyson-Schwinger equation for the vertex
    in three-gamma approximation}
  \end{center}
\end{figure}

\begin{figure}[tbp]
\begin{center}
   \resizebox{0.55\textwidth}{!}{
   \includegraphics{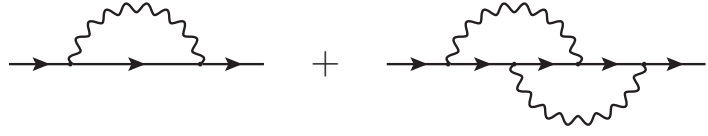}}
\caption{\label{two-loop-A}Diagrams contributing to order $1/N^2$ in the function $A(p)$.}
\end{center}
\end{figure}

We now reconsider the analysis of the fermion gap equation in the order $1/N^2$ performed by Nash following the approach described above. First, we note that in the regime when the momentum of one fermion $p$ is larger (or smaller) than the momentum $k$ of another fermion the asymptotic form of the vertex $\Gamma_\mu(p,k)$ is given by
\be
\Gamma_\mu(p,k)\simeq \Gamma_\mu(p,0)=f(p)\gamma_\mu+g(p)\left(\gamma_\mu-\frac{p_\mu\hat{p}}{p^2}\right),\quad p\gg k,
\ee
which contains only two scalar functions $f(p)$ and $g(p)$. These functions can be found solving the
DSE for the vertex in the so-called three-gamma approximation \cite{Landau} (see Fig.~\ref{DSE-vertex}).
In Appendix \ref{append-1} we derived a coupled system of equations for the functions
$f(p)$ and $g(p)$ in the above approximation and found the following solutions at order $1/N$:
\ba
f(p)&=&A(p)=\left(1+\frac{16}{9\pi^2N}\right)\left(\frac{p}{\alpha}\right)^{-2\gamma},\\
g(p)&=&D A(p),\qquad D=-\frac{8}{3\pi^2N}.
\ea
In the order $1/N^2$, the kernel $K(p,k)$ includes also the second diagram in Fig.~\ref{kernel_skeleton} [see Eq.~(\ref{K2_plus_K4}) below] and its general structure
in considered approximation can be written in the form
 \be
K(p,k)=\frac{A^2({\rm max}(p,k))}{A(p)A(k)}\tilde{K}(p,k),
 \label{K(p,k)_nota}
 \ee
where anomalous scaling functions $A$'s are factored out explicitly.
The function $A(p)$ scales as $p^{-2\gamma}$, thus we need to know the anomalous dimension at
order $1/N^2$ which can be obtained by calculating the massless fermion self-energy at two-loop level
(see Fig.~\ref{two-loop-A}) using the photon propagator (\ref{photon-propagator}).
The anomalous fermion dimension $\gamma$ at $1/N^2$ can be extracted from the Gracey's work \cite{Gracey}:
\be
\gamma=-\frac{2(2-3\xi)}{3\pi^2N}+\frac{4}{9\pi^4N^2}[64-6\pi^2+\xi(9\pi^2-92)].
\label{gamma-arbitrary-xi}
\ee
($2\gamma$ in our notations corresponds to $\eta$ in Ref. \cite{Gracey} and $-\lambda$ in Ref.~\cite{Nash}). In the gauges $\xi=0$ and $\xi=1$ we have
\ba
\gamma({\xi=0})&=&-\frac{4}{3\pi^2N}+\frac{8}{9\pi^4N^2}(32-3\pi^2),
\label{gamma-xi=0}\\
\gamma({\xi=1})&=& \frac{2}{3\pi^2N}+\frac{4}{9\pi^4N^2}(3\pi^2-28),
\label{gamma-xi=1}
\ea
respectively. As follows from Eq.~(\ref{gamma-arbitrary-xi}),  the anomalous fermion dimension vanishes
in the gauge
\be
\xi=\frac{2}{3}\left(1-\frac{8}{9\pi^2N}\right),
\label{gauge-for-zerogamma}
\ee
which is a generalization of the Nash's gauge $\xi=2/3$ to the next order in $1/N$ expansion.
This gauge is an analogue of a special gauge depending on the charge, $\xi(e_0)$, in four-dimensional
QED in which the renormalization constant $Z_2$ is finite \cite{Johnson}.

The Gracey's expression for $\gamma$ respects  the Landau-Khalatnikov-Fradkin (LKF) \cite{LK,Fradkin} transformation. Indeed, the transformation from the gauge $\xi=0$ to a gauge with arbitrary $\xi$ in Eq.~(\ref{photon-prop}) is given by
\be
S(x,\xi)=S(x,\xi=0)\exp\left[-e^2\xi\int\frac{d^3p}{(2\pi)^3}\frac{1-e^{ipx}}
{p^4[1+\Pi(p)]}\right].
\label{LKFtransform}
\ee
Calculating the integral in Eq.~(\ref{LKFtransform}) and performing the Fourier transform back to the momentum space (see Appendix~\ref{LKF-sec}), we find that at the order $1/N^2$ the anomalous fermion dimension $\gamma$ transforms as
\be
\gamma({\xi=0})\rightarrow\gamma(\xi)= \gamma({\xi=0})+\frac{2\xi}{\pi^2NC}
\label{anomdimension-transform}
\ee
with the constant $C$  defined in Eq.~(\ref{twoloop-Pi}). We note that the expression for $\lambda=-2\gamma$ ($\xi=1$) presented in Eq.~(13) of Ref.~\cite{Nash} does not agree with the expression (\ref{gamma-xi=1}) while the expressions~(\ref{gamma-xi=0}) and (\ref{gamma-xi=1}) are in agreement with LKF transformation.

In order to write down an equation for the mass function $\Sigma(p)$ we need to calculate the first
two diagrams in Fig.~\ref{kernel_skeleton}. For the contribution of the first diagram, we obtain from Eq.~(\ref{K_1_photon_exchange})
\begin{align}
K^{(2)}(p,k)&=\frac{4}{\pi^2NC{\rm max}(p,k)}\frac{1}{A(p)A(k)}
\biggl[(2+\xi)f^2({\rm max}(p,k))+4g({\rm max}(p,k))f({\rm max}(p,k))
+2g^2({\rm max}(p,k)) \nonumber \\
&\quad-\frac23(1-\xi)\frac{{\rm min}(p^2,k^2)}{{\rm max}(p^2,k^2)}g({\rm max}(p,k))
[2f({\rm max}(p,k))+g({\rm max}(p,k))]\biggr].
\end{align}
The diagram with crossed photon lines,
\begin{align}
K^{(4)}_{\beta'\alpha';\beta\alpha}(k,k,k-p)&=(ie)^4\int\frac{d^3q}{(2\pi)^3}
\bigl[\Gamma_{\lambda}(p,k+q;p-q-k)S(k+q)\Gamma_{\nu}(k+q,k;q)\bigr]_{\alpha\alpha'}
\nonumber \\
&\qquad\times\bigl[\Gamma_{\kappa}(k,p-q;k+q-p)S(p-q)\Gamma_{\mu}(p-q,p;-q)\bigr]_{\beta'\beta}
D_{\mu\nu}(q)D_{\kappa\lambda}(p-q-k),
\end{align}
gives the following contribution to the kernel:
\ba
&&K^{(4)}(p,k)=\frac{64}{N^2A(p)A(k)}\int\frac{d\Omega_k}{(2\pi)^3}\int
\frac{d^3q}{(2\pi)^3}\frac{A^2[{\rm max}(|p|,|k|,|q|)]}{|q|(p-q)^2(q+k)^2|p-q-k|}
\left\{-2(q+k)_\mu P_{\mu\nu}(q)P_{\nu\lambda}(p-q-k)\right.\nonumber\\
&&\times(p-q)_\lambda-\left.2(p-q)_\mu P_{\mu\nu}(q)P_{\nu\lambda}(p-q-k)(q+k)_\lambda
+2(p-q)_\mu(q+k)_\nu P_{\mu\nu}(q)P_{\lambda\lambda}(p-q-k)+2P_{\mu\mu}(q)\right.\nonumber\\
&&\left.\times(p-q)_\lambda(q+k)_\rho P_{\lambda\rho}(p-q-k)+(p-q)(q+k)\left[2P_{\mu\nu}(q)
P_{\nu\mu}(p-q-k)-P_{\mu\mu}(q)P_{\lambda\lambda}(p-q-k)\right]\right\}
\label{crossdiagram}
\\
&&\equiv\frac{1}{A(p)A(k)}\int\frac{d\Omega_k}{(2\pi)^3}I, \nonumber
 \ea
where we set $C=1$, $A(k+q)\simeq A(\max(k,q))$, $A(p-k)\simeq A(\max(p,k))$ and used the Landau approximation for the vertex
\begin{align}
\Gamma_\mu(k+q,k)&\simeq\gamma_\mu A(\max(k,q)),\nonumber \\
\Gamma_\mu(p-q,p)&\simeq\gamma_\mu A(\max(p,q)), \\
\Gamma_\mu(p,k+q)&\simeq\Gamma_\mu(k,p-q)\simeq\gamma_\mu A(\max(p,k,q)).\nonumber
\end{align}

Using the symmetry $K^{(4)}(p,k)=K^{(4)}(k,p)$, we evaluate it for the case $p>k$ only and then restore $p\to\max(p,k)$, $k\to\min(p,k)$ in the final result. The integrals in Eq.~(\ref{crossdiagram}) are rather difficult to calculate exactly, thus we  will approximate the $q$ integration by splitting it into two regions, $q<k$ and $q>k$, and putting the minimal momenta in each of regions equal to zero.
The used approximation  retains only the dominant asymptotics of the two-loop integral which is sufficient at order $1/N^2$ as we show below. For the first region, we find
\begin{equation}
I^{(q<k)}=\frac{16(4+8\xi+3\xi^2)A^2(p)k^2}{9\pi^4N^2p^3},
\end{equation}
and for the second region
\be
I^{(q>k)}\simeq\frac{64\xi(4+\xi)}{N^2}\int\frac{d^3q}{(2\pi)^3}
\frac{A^2[{\rm max}(p,q)](p-q)q}{|q|^3|p-q|^3}=-\frac{32\xi(4+\xi)}{\pi^2N^2(1+4\gamma)}
\frac{A^2(p)}{p},
\ee
where we integrated over the angles and then over $q$ using the explicit expression (\ref{wavefun_renorm}) for the function $A(p)$. It is interesting that this contribution vanishes in the Landau gauge $\xi=0$. Thus the diagram with crossed photon lines gives
\be
K^{(4)}(p,k) =\biggl[-\frac{16\xi(4+\xi)}{\pi^4N^2(1+4\gamma)}
+\frac{8(4+8\xi+3\xi^2)}{9\pi^4N^2}\frac{{\rm min}(p^2,k^2)}{{\rm max}(p^2,k^2)}\biggr]\frac{1}{{\rm max}(p,k)}
\frac{A^2[{\rm max}(p,k)]}{A(p)A(k)}.
\label{kernel(p>k)}
\ee
Combining
Eq.~(\ref{kernel(p>k)}) with the contribution of first diagram in Fig.~\ref{kernel_skeleton} we
finally find the expression for the kernel:
\ba
K(p,k)\simeq\left[\frac{4(2+\xi)}{\pi^2N}\left(1-\frac{c}{\pi^2N}\right)+
\frac{8(20-8\xi+3\xi^2)}{9\pi^4N^2}
\frac{{\rm min}(p^2,k^2)}{{\rm max}(p^2,k^2)}\right]
\frac{1}{{\rm max}(p,k)}\frac{A^2[{\rm max}(p,k)]}{A(p)A(k)}
\label{K2_plus_K4}
\ea
with the constant $c=a+[12\xi(4+\xi)+32]/3(2+\xi)$. In the above expression we kept only terms of order $1/N^2$. In the gauge $\xi=1$, $c=a+92/9$ which is different from the value $c=(80+6a)/9$ given in Ref. \cite{Nash}.

Equation~(\ref{eq-for-b-exponent})  then yields the following equation for the exponent $b$:
\begin{equation}
1=\frac{4(2+\xi)}{\pi^2N}\biggl(1-\frac{c}{\pi^2N}\biggr)\frac{1+4\gamma}{b(1-b)+2\gamma(1+2\gamma)}
+\frac{8(20-8\xi+3\xi^2)}{9\pi^4N^2}\frac{5+4\gamma}{b(1-b)+2(\gamma+1)(3+2\gamma)},
\label{eq_for_b}
\end{equation}
or, keeping only the terms up to the order $1/N^2$,
\be
b(1-b)=\frac{32}{3\pi^2N}+\frac{64(3\pi^2-44)}{9\pi^4N^2}=
\frac{32}{3\pi^2N}\left(1-\frac{9.59}{\pi^2N}\right).
\label{gauge_inv_b:eq}
\ee
Note that the second term in Eq.~(\ref{eq_for_b}), which originates from the terms proportional to $\min(p^2,k^2)/\max(p^2,k^2)$ in the kernel $K(p,k)$, does not contribute to Eq.~(\ref{gauge_inv_b:eq})
in the order $1/N^2$.
It is remarkable that the dependence on the gauge parameter $\xi$ has completely cancelled out so
that the exponent $b$ is indeed gauge independent. From  the equation for $b$ we find the critical
$N_c=2.85$ which should be contrasted to the value $N_c=3.28$ found by Nash.
We can compute from Eq.~(\ref{gauge_inv_b:eq}) the mass anomalous dimension $\gamma_m=1-b$ in $1/N$ expansion,
\be
\gamma_m\simeq
\frac{32}{3\pi^2N}+\frac{64(3\pi^2-28)}{9\pi^4N^2},
\ee
which coincides with that found by Gracey \cite{Gracey} (our definition of $\gamma_m$ has the sign
opposite to Gracey's $\gamma_m$).

\section{Estimate of the chiral condensate}
\label{chi-condensate}
To get an estimate of the chiral condensate, which is a gauge independent quantity, we now
turn to studying Eq.~(\ref{main:eq}) and consider it in a gauge where the anomalous dimension $\gamma$ vanishes; i.e., we take the gauge parameter $\xi$ as given in Eq.~(\ref{gauge-for-zerogamma}). Then Eq.~(\ref{main:eq}) considerably simplifies and takes the form
\be
\Sigma(p)=\lambda\int\limits_0^\alpha\frac{dk\,k^2\Sigma(k)}{k^2+\Sigma^2(k)}
\frac{1}{{\rm max}(p,k)},\quad\lambda=\frac{32}{3\pi^2N}+\frac{64(3\pi^2-44)}{9\pi^4N^2}
\label{integeq-Sigma}
\ee
(since the region $k<\Sigma(k)$ gives a negligible contribution to the integral, see e.g. Ref.~\cite{Nash}, one can still use the kernel $K(p,k)$ calculated in the massless theory).
The above integral equation is equivalent to the following differential equation
\be
(p^2\Sigma^\prime(p))^\prime=-\lambda\frac{p^2\Sigma(p)}{p^2+\Sigma^2(p)},
\label{diff-eq}
\ee
with the IR and ultraviolet (UV) boundary conditions
\be
p^2\Sigma^\prime(p)\Bigr|_{p=0}=0,\quad (p\Sigma(p))^\prime\Bigr|_{p=\alpha}=0.
\label{BC}
\ee
For the chiral condensate we get
\be
\langle\bar\psi\psi\rangle=-4\int\frac{d^3p}{(2\pi)^3}\frac{\Sigma(p)}{p^2+\Sigma^2(p)}
=-\frac{2}{\pi^2}\int\limits_0^\alpha\frac{dp\,p^2\Sigma(p)}{p^2+\Sigma^2(p)}
=\frac{2\alpha^2}{\pi^2\lambda}\Sigma^\prime(p)\Bigr|_{p=\alpha}
=-\frac{2\alpha}{\pi^2\lambda}\Sigma(p=\alpha),
\ee
where in the last two equalities we used the differential equation (\ref{diff-eq}) and the
UV boundary condition~(\ref{BC}).
Note that we do not include the factor $N$ (summation over flavors) in the definition
of the condensate. The mass function has the following asymptotics for momenta $p\gg\Sigma_0$:
\be
\Sigma(p)=A_0\frac{\Sigma_0^{3/2}}{\nu\sqrt{p}}\sin\left[\frac{\nu}{2}\left(
\ln\frac{p}{\Sigma_0}+\delta\right)\right],\quad \nu=\sqrt{4\lambda-1},
\label{Sigma-asympt}
\ee
with $\Sigma_0$ being the overall scale of the solution $\Sigma(p)$, $A_0$ some constant of order
one and $\delta$ is a phase.

\begin{figure}[h]
\includegraphics[width=7cm]{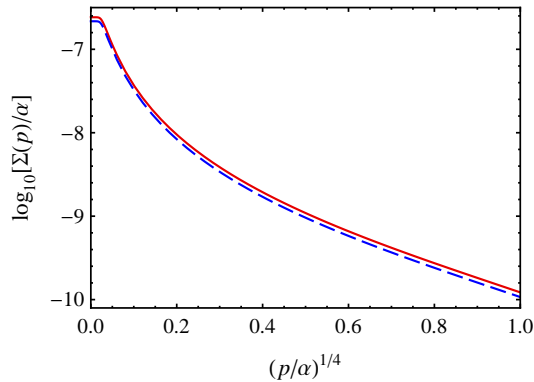}
\caption{Numerical solution of Eq.~(\ref{integeq-Sigma}) (solid line) and an approximate analytical solution~(\ref{Sigma_approx}) (dashed line) at $N=2$ plotted on a logarithmic scale.}
\label{Sigma_N2}
\end{figure}

The UV boundary condition~(\ref{BC}) leads to the following solution for the scale $\Sigma_0$:
\be
\Sigma_0=\alpha\exp\left(-\frac{2\pi}{\nu}+\delta+\frac{2\tan^{-1}\nu}{\nu}\right).
\ee
Then, for the dimensionless condensate, we get
\be
\frac{\langle\bar\psi\psi\rangle}{e^4}=\frac{N^2\langle\bar\psi\psi\rangle}{64\alpha^2}
=-\frac{N^2A_0}{128\pi^2\lambda^2}\left(\frac{\Sigma_0}{\alpha}\right)^{3/2}.
\ee
To get estimates of values $A_0$ and $\Sigma_0/\alpha$ we first use the solution
of the linearized equation (\ref{integeq-Sigma}) when $\Sigma^2(p)$ in the denominator
is replaced by $\Sigma^2(0)\equiv\Sigma_0^2$. Then the solution is
\be
\Sigma(p)=\Sigma_0F\left(\frac{1+i\nu}{4},\frac{1-i\nu}{4};\frac{3}{2};
-\frac{p^2}{\Sigma^2_0}\right).
\label{Sigma_approx}
\ee
Its asymptotics has the form of Eq.~(\ref{Sigma-asympt}) with
\be
A_0=2\sqrt{\pi}\Big|\frac{\Gamma\left(1+\frac{i\nu}{2}\right)}{\Gamma
\left(\frac{5+i\nu}{4}\right)\Gamma\left(\frac{1+i\nu}{4}\right)}\Big|,
\quad \delta=\frac{2}{\nu}\,{\rm Arg}\left[\frac{\Gamma\left(1+\frac{i\nu}{2}
\right)}{\Gamma
\left(\frac{5+i\nu}{4}\right)\Gamma\left(\frac{1+i\nu}{4}\right)}\right].
\ee
For $N=2$, we find
\be
A_0\simeq1.12, \quad \delta\simeq 1.59,\quad \frac{\Sigma_0}{\alpha}\simeq
2.17\times 10^{-7};
\ee
hence we get the estimate for the condensate
\be
\frac{\langle\bar\psi\psi\rangle}{e^4}\approx -4.64\times10^{-12}.
\label{cond-estimate}
\ee
Solving Eq.~(\ref{integeq-Sigma}) numerically for $N=2$ (see Fig.~\ref{Sigma_N2}), we get
\begin{equation}
\frac{\Sigma(0)}\alpha\approx2.42\times10^{-7}, \qquad
\frac{\langle\bar\psi\psi\rangle}{e^4}\approx-5.57\times10^{-12},
\end{equation}
which is very close to the estimate (\ref{cond-estimate}). As is seen, the condensate is very small
for $N=2$ in order to be extracted from lattice simulations \cite{lattice}. Also, the smallness
of the quantity $\Sigma_0/\alpha$ justifies the neglect of the terms quadratic or higher in the mass function $\Sigma(k)$ in the kernel (\ref{K(p,k)_notation}). Obviously, the smallness of both quantities,
$\langle\bar\psi\psi\rangle/e^4$ and $\Sigma_0/\alpha$, is due to the proximity of the fermion
number $N=2$ to the critical value $N_c$.

\section{Conclusion}
\label{section-conclusion}
In the present paper, we reconsidered an analysis of calculating the critical number $N_c$
in three-dimensional QED with $N$ four-component massless fermions. The value $N_c$ marks the quantum critical point separating chiral symmetric and chiral symmetry broken phases. While at present the
majority of works  agree on the existence of $N_c$, its precise value remains a matter of debate.
Our analysis of the Dyson-Schwinger equation for the fermion dynamical mass uses an approach advocated
by Holdom \cite{Holdom} and Mahanta \cite{Mahanta} in case of QED4; it deviates in some important
details from Nash's approach \cite{Nash}.

We obtained the critical value $N_c\approx2.85$ which differs
from the value $N_c\approx3.28$ claimed by Nash and the value $N_c\approx3.29$ obtained in the recent
paper \cite{Kotikov2016}. More importantly, we show that our $N_c$ is gauge independent in contrast
to Refs.\cite{Nash,Kotikov2016}. Our critical $N_c\approx2.85$ means that the chiral symmetry 
is dynamically broken for integer values $N \leq2$, while for $N\ge 3$ the system is in a chirally symmetric phase. The value $N=2$ is also well inside the bound $N<N_c=1+\sqrt{2}\approx 2.414$ obtained in a conformal QED3 \cite{Giombi_JHEP2016}.

The fermion mass anomalous dimension $\gamma_m$ calculated in our
approach turns to be also gauge independent and agrees with the one obtained by Gracey  to the
order $1/N^2$ \cite{Gracey}. The form used by us for the wave-function renormalization is in agreement
with perturbative calculations in the $1/N$ order and at the same time satisfies the Landau-Khalatnikov-Fradkin transformation rules for the transition between different covariant gauges.
Our numerical estimate of  chiral condensate gives a rather small value for $N=2$, due to its proximity  to the critical value $N_c$,
which explains difficulties for extracting the condensate from current lattice simulations.

\begin{acknowledgments}
We are grateful to Anatoly Kotikov and Sofian Teber for discussions and information about their work
in this direction.
The work of V.P.G. is supported by the Program of Fundamental Research of the Physics and Astronomy
Division of the National Academy of Sciences of Ukraine and by the RISE Project CoExAN No. GA644076.
\end{acknowledgments}

{\it Note added}.---Recently, Ref.~\cite{Herbut} appeared,
where the critical value $N_c\approx2.89$ in QED3 was obtained using completely different approach,
namely, from a condition of an annihilation of an infrared-stable fixed point that describes the
large-$N$ conformal phase by another unstable fixed point at a critical number of fermions $N = N_c$.

\appendix

\section{One-loop calculation of the  wave-function renormalization.}
\label{A-function}
For the fermion wave-function renormalization $A(p)$, we have the following expression in the one-loop
approximation in perturbation theory:
\ba
A(p)=1+\frac{e^2}{p^2}\int\frac{d^3q}{(2\pi)^3}\frac{1}{q^2(q^2+\alpha|q|)}
\frac{\xi p^2q^2+(2-\xi)q^2(pq)+2(1-\xi)(pq)^2}{(p+q)^2}.
\label{1loop_A-function}
\ea
Usually, since we are interested in $1/N$ expansion, we neglect the term
$q^2$ compared with $\alpha|q|$ in the denominator of the last expression.
 In this case, we actually deal with
renormalizable QED$3$ with the photon propagator $\sim1/|q|$. The fermion
self-energy (or the function $A$) becomes linearly (superficially) divergent so that
the integral over $q$ requires a cutoff which is taken at $\alpha$.
However, because of a linear divergence of the self-energy, there is ambiguity
in the constant term. Indeed, if $q$ is the photon momentum of integration,
then
\be
A(p)=1+\frac{4}{\pi^2N}\left[\left(\xi-\frac{2}{3}\right)\ln\frac{\alpha}{p}+
\frac{4}{9}\right].
\label{right-A-inPT}
\ee
On the other hand, if $q$ is the fermion momentum of integration, one gets
\be
A(p)=1+\frac{4(3\xi-2)}{\pi^2N}\left(\ln\frac{\alpha}{p}+
\frac{1}{3}\right).
\ee
This ambiguity can be easily solved if we start from the expression~(\ref{1loop_A-function})
in full QED$3$ with the photon propagator given by Eqs.~(\ref{photon-prop}) and (\ref{twoloop-Pi}).
In this case, the self-energy is only (superficially) logarithmically divergent and the result
does not depend on the choice of integration momentum. Moreover, the expression for the $A$
function turns out to be finite after angular integration and takes the form
\ba
A(p)=1+\frac{2}{\pi^2N}\frac{\alpha}{p}\int\limits_0^\infty\frac{dx}{x(x+\alpha/p)}
f(x)=1+\frac{2}{\pi^2N}\int\limits_0^\infty dx\left(\frac{1}{x}-\frac{1}{x+\alpha/p}
\right)f(x),
\label{full-1loop-A}
\ea
where
\be
f(x)=(x^2-1)\left(1-\frac{x^2+1}{4x}L\right)+\xi\left(1+\frac{x^2-1}{4x}L\right),\quad
L=\ln\left(\frac{1+x}{1-x}\right)^2.
\ee
The integral in Eq.~(\ref{full-1loop-A}) is convergent, and to extract its behavior at large $\alpha/p$,
we first write
\ba
\int\limits_0^\infty dx\left(\frac{1}{x}-\frac{1}{x+\alpha/p}\right)f(x)
&=&\int\limits_0^1 dx\left(\frac{1}{x}-\frac{1}{x+\alpha/p}\right)f(x)+\int\limits_1^\infty dx\left(\frac{1}{x}-\frac{1}{x+\alpha/p}\right)[f(x)+\frac{4}{3}-2\xi]\nonumber\\
&+&\left(2\xi-\frac{4}{3}\right)\int\limits_1^\infty dx\left(\frac{1}{x}-\frac{1}{x+\alpha/p}\right).
\ea
In the first two terms in the right-hand side of the last equation, we take the limit $\alpha\to\infty$, and
while the third term is explicitly calculated, we get
\ba
\int\limits_0^\infty dx\left(\frac{1}{x}-\frac{1}{x+\alpha/p}\right)f(x)&\simeq&
\biggl[\frac{4}{3}(1-\ln2)-\xi(1-2\ln2)\biggr]+\biggl[-\frac{4}{9}(1-3\ln2)+\xi(1-2\ln2)\biggr]+
\frac{2(3\xi-2)}{3}\ln\frac{\alpha}{p}\nonumber\\
&=&\frac{2(3\xi-2)}{3}\ln\frac{\alpha}{p}+\frac{8}{9}.
\ea
Thus we obtain
\be
A(p)=1+\frac{4(3\xi-2)}{3\pi^2N}\ln\frac{\alpha}{p}+\frac{16}{9\pi^2N}.
\ee
From the last expression we find the fermion anomalous dimension at the $1/N$ order
\be
\gamma=-\frac{1}{2}\,p\frac{d\ln A(p)}{d p}= \frac{2(3\xi-2)}{3\pi^2N},
\ee
and at the $1/N^2$ order it was calculated by Gracey \cite{Gracey} (see Eq.~(\ref{gamma-arbitrary-xi})).
We obtain the expression given by Eq.~({\ref{right-A-inPT}}) which can be written
approximately as
\be
A(p)\simeq\left(1+\frac{16}{9\pi^2N}\right)\left(\frac{p}{\alpha}\right)^{-2\gamma}.
\ee
One can check that the coefficient $a=1+16/9\pi^2N$ before the power in last
expression is in agreement with corresponding findings of Gracey \cite{Gracey}.
 Note that $A(p)$ is not
identically equal to $1$ in the gauge $\xi=2/3$, the difference is in a constant term.

\section{LKF transformation}
\label{LKF-sec}
The LKF transformation relating the coordinate-space fermion propagator in two different
 covariant gauges $\xi$ and $\xi'$ follows from Eq.~(\ref{LKFtransform}),
\be
S(x,\xi)=S(x,\xi')\exp\left[-e^2(\xi-\xi')\int\frac{d^3p}{(2\pi)^3}\frac{1-e^{ipx}}{p^4[1+\Pi(p)]}
\right]\equiv S(x,\xi')G(x,\xi-\xi').
\label{LKFtransform1}
\ee
The LKF transformation for the vertex is more complicated and can be found in Ref. \cite{Landau}
(see also \cite{Burden}). The DS equations as well as WT identities are covariant under the LKF
transformation in configuration space but corresponding transformations look much more complicated
in momentum space.

 Using the polarization function with massless fermions in two loops (\ref{twoloop-Pi}) the integral
in Eq.~(\ref{LKFtransform1}) can be evaluated exactly,
\ba
&&e^2(\xi-\xi')\int\frac{d^3p}{(2\pi)^3}\frac{1-e^{ipx}}{p^4[1+\Pi(p)]}=\frac{4d}{b\pi}
\left\{{\log(b|x|)+\gamma-1}+\frac{\pi}{b|x|}\sin^2\frac{b|x|}{2}\right.\nonumber\\
&&+\left.\frac{1}{b|x|}\left[\cos(b|x|)\mathrm{Si}(b|x|)-\sin(b|x|)\mathrm{Ci}(b|x|)\right]\right\}
\equiv \frac{d}{b}f(b|x|),\quad b=\frac{e^2NC}{8},\quad d=\frac{e^2(\xi-\xi')}{8\pi},
\ea
where $\mathrm{Si}(z)$ and $\mathrm{Ci}(z)$ are the sine and cosine integral functions, respectively, and $\gamma$ is the Euler constant (not to be confused with the fermion anomalous dimension). We find the following asymptotics:
\ba
\frac{d}{b}f(b|x|)\simeq\left\{\begin{array}{cl}d|x|,\quad & b|x|\ll1,\\
\frac{4d}{\pi b}\log\left(e^{\gamma-1}b|x|\right),\quad& b|x|\gg1. \end{array}\right.
\ea
In the $1/N$ expansion, only the asymptotics at $b|x|\gg1$ is important, leading to the transformation
rule~(\ref{anomdimension-transform}) for the anomalous dimension.

Performing the Fourier transform in Eq.~(\ref{LKFtransform1}) we obtain the transformations
in momentum space relating the wave-function renormalization and the mass function in two different
gauges:
\ba
\frac{1}{A(p;\xi)[p^2+\Sigma^2(p;\xi)]}&=&\frac{1}{p^2}\int\frac{d^3k}{(2\pi)^3}
\frac{(pk)}{A(k;\xi')[k^2+\Sigma^2(k;\xi')]}G(p-k,\xi-\xi'),
\label{transform-nonlocal:F}\\
\frac{\Sigma(p;\xi)}{A(p;\xi)[p^2+\Sigma^2(p;\xi)]}&=&
\int\frac{d^3k}{(2\pi)^3}\frac{\Sigma(k;\xi')}{A(k;\xi')[k^2+\Sigma^2(k;\xi')]}G(p-k,\xi-\xi'),
\label{mass-nonlocal:eq}
\ea
where $G(p,\xi-\xi')$ is the Fourier transform of $G(|x|,\xi-\xi')$. The limit
$\lim_{\xi\to\xi'_+}G(p,\xi-\xi')=(2\pi)^3\delta^{(3)}(p)$ confirms the self-consistency of the momentum-space LKF transform. Also, although the above considerations were constrained to the values $\xi>\xi'$, the transition to the region $\xi<\xi'$ is straightforward because the expansion in $\xi$
is analytic at $\xi'$.

Clearly, if the dynamical mass  $\Sigma(k;\xi')$ vanishes in the $\xi'$-gauge at some critical $N_c$,
then it vanishes in an arbitrary gauge $\xi$, which  means that the value $N_c$ is gauge independent.
Certainly, the covariance of the fermion propagator is crucial for a gauge independence of $N_c$.

To evaluate the condensate
\be
\langle\bar\psi\psi\rangle=-{\rm tr}S(x=0,\xi)=-4\int \frac{d^3p}{(2\pi)^3}\frac{\Sigma(p;\xi)}{A(p;\xi)[p^2+\Sigma^2(p;\xi)]}
\ee
we should integrate Eq.~(\ref{mass-nonlocal:eq}) over $\int d^3p/(2\pi)^3$. Using the fact that
\be
\int \frac{d^3p}{(2\pi)^3}G(p,\xi-\xi')=G(x=0,\xi-\xi')=1,
\ee
we establish the gauge independence of the chiral condensate.

For massless case, $\Sigma=0$, starting from the gauge (\ref{gauge-for-zerogamma}) where the wave-function renormalization $A(p)\simeq \text{const}$, Eq.~(\ref{transform-nonlocal:F}) allows us to obtain
$A(p)$ in a gauge with an arbitrary gauge parameter $\xi$:
\be
A(p)\simeq \text{const}\left(\frac{p}{\alpha}\right)^{-2\gamma},
\ee
where $\gamma$ is given by Eq.~(\ref{gamma-arbitrary-xi}).

\section{The DS equation for the vertex in three-gamma approximation}
\label{append-1}

The Dyson-Schwinger equation for the vertex in the Landau (three-gamma) approximation has the form:
\ba
\Gamma_\mu(p,k;p-k)=\gamma_\mu+(ie)^2\int\frac{d^3q}{(2\pi)^3}\Gamma_\lambda
(p,p+q;q)S(p+q)\Gamma_\mu(p+q,k+q;p-k)S(k+q)\Gamma_\rho(k+q,k;q)D_{\lambda\rho}(q),
\label{vetex:eq}
\ea
where the photon propagator $D_{\lambda\rho}(q)$ is taken in the leading $1/N$
approximation~(\ref{photon-prop}) with the vacuum polarization from Eq.~(\ref{twoloop-Pi}).

From the DS equation (\ref{fermion:eq}) for the full fermion propagator
 we get the equation for the function $A(p)$:
\be
A(p)=1+\frac{e^2}{p^2}\int\frac{d^3q}{(2\pi)^3}\frac{{\rm tr}[\hat{p}\gamma_\mu(\hat{p}+
\hat{q})\Gamma_\nu(p+q,p;q)]}{(p+q)^2A(p+q)}D_{\mu\nu}(q).
\label{eq:A}
\ee
Eqs.~(\ref{vetex:eq}), (\ref{photon:eq}), (\ref{fermion:eq}) represent the system of equations
for the vertex, the photon and fermion propagators in massless theory truncated in $1/N$
approximation and should be solved in nonperturbative way. They  lead to a system of equations
for $A(p)$ and some scalar functions before tensor structures in $\Gamma_\mu$. The gauge
invariance requires the vertex to satisfy the Ward-Takahashi identity (\ref{vectorWTI}).
We will study the equation for the vertex when one of fermion momenta is much larger than
the other, $|p|\gg |k|$. Setting $k=0$ in Eq.~(\ref{vetex:eq}) we obtain
\ba
\Gamma_\mu(p,0;p)=\gamma_\mu+\frac{8}{NC}\int\frac{d^3q}{(2\pi)^3}
\frac{1}{|q|^3(p+q)^2A(p+q)A(q)}\Gamma_\lambda
(p,p+q;q)(\hat{p}+\hat{q})\Gamma_\mu(p+q,q;p)\hat{q}\Gamma_\rho(q,0;q)P_{\lambda\rho}(q).
\ea
For $\Gamma_\mu(p,0;p)$ we use
\be
\Gamma_\mu(p,0;p)=f(p)\gamma_\mu+g(p)\left(\gamma_\mu-\frac{p_\mu\hat{p}}{p^2}\right),\quad |p|\gg |k|,
\ee
which is in agreement with the WT identity. A similar expression is valid for $\Gamma_\mu(0,p;p)$.
Actually, the WT identity requires that $f(p)=A(p)$ for massless fermions
and we will see that solutions of our system of equations really satisfy this condition.

For other vertices we take an approximation
\ba
\Gamma_\lambda(p,p+q;q)&\simeq&\Gamma_\lambda(p,p;0)\theta(p-q)+\Gamma_\lambda(0,q;q)\theta(q-p),\\
\Gamma_\mu(p+q,q;p)&\simeq&\Gamma_\mu(p,0;p)\theta(p-q)+\Gamma_\mu(q,q;0)\theta(q-p).
\ea
For the vertex with zero photon momentum we can use the WTI
\be
\Gamma_\mu(q,q;0)=i\frac{\partial S^{-1}(q)}{\partial q_\mu}\simeq \gamma_\mu A(q),
\ee
where we neglected the derivative of the $A$ function which is of higher order in $1/N$. Similarly,
for the $A$ function we use the approximation
$A(p+q)\simeq A(\mbox{max}(p,q))=A(p)\theta(p-q)+A(q)\theta(q-p)$.
The above made approximations thus assume that we can neglect an angular dependence
in dimensionless scalar functions.

After some algebraic work we get a coupled system of equations for scalar functions
$f(p)$ and $g(p)$:
\ba
f(p)&=&1+\frac{8}{NCp^2}\int\frac{d^3q}{(2\pi)^3}\frac{1}{|q|^3(p+q)^2A(q)}
\left\{f(p)\left[f(q)\left(-2p^2pq-2(pq)^2+\xi p^2q(p+q)\right)\right.\right.\nonumber\\
&-&\left.\left.2pq p(p+q)g(q)\right]\theta(p^2-q^2)+\left[f^2(q)\left(-2p^2pq-2(pq)^2
+\xi p^2q(p+q)\right)\right.\right.\nonumber\\
&-&\left.\left.2pq p(p+q)g(q)[2f(q)+g(q)]\right]\theta(q^2-p^2)\right\},
\label{f-function:eq}\\
f(p)+\frac{2}{3}g(p)&=&1+\frac{8}{NC}\int\frac{d^3q}{(2\pi)^3}
\frac{1}{|q|^3(p+q)^2A(q)}\left\{\left[[f(p)+g(p)][(\xi-\frac{2}{3})f(q)-
\frac{2}{3}g(q)]q(p+q)\right.\right.\nonumber\\
&+&\left.\left.\frac{1}{3}g(p)f(q)\left(2pq+2\frac{(pq)^2}{p^2}
-\xi q(p+q)\right)+\frac{2}{3}g(p)g(q)(pq+\frac{(pq)^2}{p^2})\right]
\theta(p^2-q^2)\right.\nonumber\\
&+&\left.\left[(\xi-\frac{2}{3})f^2(q)-
\frac{2}{3}g(q)[2f(q)+g(q)]\right]q(p+q)\theta(q^2-p^2)\right\}
\ea
Subtracting the first equation from the second one, we see that the function $g(p^2)$
is of order $1/N$, thus we can neglect in resulting equations all functions
$g$'s in the integrand, and the equations for $f,g$ functions become decoupled:
\ba
f(p)&\simeq&1+\frac{8}{NC}\int\frac{d^3q}{(2\pi)^3}\frac{1}{|q|^3(p+q)^2A(q)}
\left[{f(p)f(q)}\theta(p^2-q^2)+f^2(q)\theta(q^2-p^2)\right]\nonumber\\
&\times &\left(-2(pq)-2\frac{(pq)^2}{p^2}+\xi q(p+q)\right),\\
g(p)&\simeq&\frac{24}{NC}\int\frac{d^3q}{(2\pi)^3}\frac{1}{|q|^3(p+q)^2
A(q)}\left[f(p)f(q)\theta(p^2-q^2)+f^2(q)\theta(q^2-p^2)\right]\nonumber\\
&\times &\left(\frac{2}{3}(pq)-\frac{1}{3}q^2+\frac{(pq)^2}{p^2}\right).
\ea
The angular integration can be performed by means of integrals:
\ba
\int\frac{d\Omega_q}{4\pi}\frac{1}{(q+p)^2}&=&\frac{1}{4pq}\ln\left(\frac{p+q}
{p-q}\right)^2,\\
\int\frac{d\Omega_q}{4\pi}\frac{pq}{(q+p)^2}&=&\frac{1}{2}\left[1-\frac{p^2
+q^2}{4pq}\ln\left(\frac{p+q}{p-q}\right)^2\right],\\
\int\frac{d\Omega_q}{4\pi}\frac{(pq)^2}{(q+p)^2}&=&\frac{p^2+q^2}{4}
\left[\frac{p^2+q^2}{4pq}\ln\left(\frac{p+q}{p-q}\right)^2-1\right].
\ea
In the right-hand side of the last equations $p\equiv|p|,q\equiv|q|$.
Thus equations for $f,g$ functions take the form
\ba
f(p)&=&1+\frac{2}{\pi^2NC}\int\limits_0^{\alpha}\frac{dq}{q}\frac{1}{A(q)}
\left[{f(p)f(q)}\theta(p-q)+f^2(q)\theta(q-p)\right]\nonumber\\
&\times&\left[\frac{q^2-p^2}
{p^2}\left(1-\frac{p^2+q^2}{4pq}L\right)+\xi\left(1+\frac{q^2-p^2}
{4pq}L\right)\right],
\label{f:eq}\\
g(p)&=&\frac{1}{\pi^2NC}\int\limits_0^{\alpha}\frac{dq}{q}\frac{1}{A(q)}
\left[{f(p)f(q)}\theta(p-q)+f^2(q)\theta(q-p)\right]\nonumber\\
&\times&\left[1-\frac{3q^2}{p^2}+\frac{(3q^2+p^2)(q^2-p^2)}{4p^3q}
L\right], \quad L=\ln\left(\frac{p+q}{p-q}\right)^2.
\label{g:eq}
\ea
First, we solve an equation for $f(p)$ which we seek in the form when
$f(p)$ is proportional to the function $A(p)$, $f(p)=E A(p)$. Plugging this in
Eq.~(\ref{f:eq}) and making the change of a variable $q=xp$, we get
\ba
E A(p)=1&+&\frac{2E^2}{\pi^2NC}A(p)\left\{\int\limits_0^1\frac{dx}{x}\left[(x^2-1)
\left(1-\frac{x^2+1}{4x}L\right)+\xi\left(1+\frac{x^2-1}{4x}L\right)\right]
\right.\nonumber\\
&+&\left.\int\limits_1^{\alpha/p}\frac{dx}{x}x^{-2\gamma}\left[(x^2-1)
\left(1-\frac{x^2+1}{4x}L\right)+\frac{4}{3}+\xi\left(-1+\frac{x^2-1}{4x}L
\right)\right]\right.\nonumber\\
&+&\left.2(\xi-\frac{2}{3})\int\limits_1^{\alpha/p}
\frac{dx}{x}x^{-2\gamma}\right\}.
\ea
In the above equation, we used the explicit form (\ref{wavefun_renorm}) of the $A(p)$ function.
Since we need the function $f(p)$ at the order $1/N$, we can set $C\simeq1$ and take $\gamma=0$
in the second integral as well as put the upper limit of integration in the second integral equal
to infinity. Then all integrals are exactly computed,
\ba
&&\int\limits_0^1\frac{dx}{x}\left[(x^2-1)
\left(1-\frac{x^2+1}{4x}L\right)+\xi\left(1+\frac{x^2-1}{4x}L\right)\right]=
\frac{4}{3}(1-\ln2)+\xi(-1+2\ln2),\\
&&\int\limits_1^{\infty}\frac{dx}{x}\left[(x^2-1)
\left(1-\frac{x^2+1}{4x}L\right)+\frac{4}{3}+\xi\left(-1+\frac{x^2-1}{4x}L
\right)\right]=\frac{4}{9}(-1+3\ln2)+\xi(1-2\ln2).
\ea
We get the following equation:
\be
E A(p)=1+\frac{2E^2}{\pi^2N}A(p)\left[\frac{8}{9}+\frac{\xi-2/3}{\gamma}
\left(1-\left(\frac{\alpha}{p}\right)^{-2\gamma}\right)\right].
\ee
Using expression (\ref{wavefun_renorm}) for $A(p)$ in the above equation and comparing terms with and without $(\alpha/p)^{-2\gamma}$,  we find
\ba
\gamma&=&\frac{2E^2\kappa(\xi-2/3)}{\pi^2N}, \qquad \kappa\equiv1+\frac{16}{9\pi^2N},\\
E&=&\frac{2E^2}{\pi^2N}\left[\frac{8}{9}+\frac{\xi-2/3}{\gamma}\right]=
\frac{16E^2}{9\pi^2N}+\frac{1}{\kappa}\simeq1+\frac{16}{9\pi^2N}(E^2-1).
\ea
This is identically satisfied if we take $E=1$; hence $f(p)=A(p)$ which is consistent with
WTI up to the order $1/N^2$. Plugging this into Eq.~(\ref{g:eq}) and seeking the solution in
the form $g(p)=DA(p)$, we obtain
\ba
g(p)=DA(p)=\frac{E^2}{\pi^2NC}\int\limits_0^{\alpha}\frac{dq}{q}\left[{A(p)}\theta(p-q)+A(q)
\theta(q-p)\right]\left[1-\frac{3q^2}{p^2}+\frac{(3q^2+p^2)(q^2-p^2)}{4p^3q}L\right].
\ea
Again, using the explicit form of the $A(p)$ function and making the change of the variable $q=xp$, we find
at order $1/N$ that
\be
D\simeq\frac{1}{\pi^2N}\int\limits_0^\infty\frac{dx}{x}\left[1-3x^2+
\frac{(3x^2+1)(x^2-1)}{4x}L\right]=-\frac{8}{3\pi^2N}.
\ee

\end{document}